\documentclass[journal,onecolumn,12pt,twoside]{IEEEtran}
\ifCLASSINFOpdf
   \usepackage[pdftex]{graphicx}
\else
   \usepackage[dvips]{graphicx}
\fi
\usepackage[cmex10]{amsmath}
\usepackage{algorithm}
\usepackage{algorithmic}
\usepackage{amsfonts, amssymb, amscd, amsthm, xspace}

\usepackage{color}
\usepackage{graphicx}
\usepackage{epstopdf}

\newtheorem{theorem}{Theorem}
\newtheorem{lemma}{Lemma}
\newtheorem{claim}{Claim}

\newtheorem{remark}{Remark}
\newtheorem{construction}{Construction}

\newcommand{\etal}{{\it et al.}}
\newcommand{\ie}{{\it i.e.}}

\newcommand{\rank}{\textrm{rank}}
\newcommand{\twomatrix}[2]{\begin{bmatrix} #1 \\ #2\end{bmatrix}}

\newcommand{\Sf}{S} 
\newcommand{\X}{X} 
\newcommand{\al}{\alpha} 
\newcommand{\be}{\beta} 
\newcommand{\GF}[1]{\mathbb{F}_{#1}} 
\newcommand{\E}{E} 
\newcommand{\G}{\mathcal{G}} 
\newcommand{\Gprime}{\mathcal{G}'} 
\newcommand{\SG}{S_{\mathcal{G}}} 
\newcommand{\SGprime}{S_{\Gprime}} 
\newcommand{\Bs}{B_s} 
\newcommand{\lam}{\lambda}
\newcommand{\lamp}{\lambda'}
\newcommand{\lams}[1]{\lambda_{#1}}
\newcommand{\gam}{\gamma}

\newcommand{\Psih}{\hat{\Psi}}
\newcommand{\Psit}{\tilde{\Psi}}
\newcommand{\Ind}[1]{\mathcal{I}\left(#1\right)} 

\newcommand{\I}[2]{I\left(#1;#2\right)} 
\newcommand{\code}{\mathcal{C}} 

\newcommand\blfootnote[1]{%
  \begingroup
  \renewcommand\thefootnote{}\footnote{#1}%
  \addtocounter{footnote}{-1}%
  \endgroup
}

\renewcommand{\baselinestretch}{0.959}

\begin{document}
%
\title{Weakly Secure Regenerating Codes for Distributed Storage}

\author{\IEEEauthorblockN{Swanand Kadhe and Alex Sprintson}}


%


\maketitle

\begin{abstract}
We consider the problem of secure distributed data storage under the paradigm of \emph{weak security}, in which no \emph{meaningful information} is leaked to the eavesdropper. More specifically, the eavesdropper cannot get any information about any individual message file or a small group of files. The key benefit of the weak security paradigm is that it incurs no loss in the storage capacity, which makes it practically appealing. 

In this paper, we present a coding scheme, using a coset coding based outer code and a Product-Matrix Minimum Bandwidth Regenerating code (proposed by Rashmi et al.) as an inner code, that achieves weak security when the eavesdropper can observe any single storage node.  We show that the proposed construction has good security properties and requires small finite field size.
\end{abstract}


\blfootnote{Swanand Kadhe and Alex Sprintson are with the Department of Electrical and Computer Engineering at Texas A\&M University, USA; Emails:\{kswanand1,spalex\}@tamu.edu.}

%
\IEEEpeerreviewmaketitle

\section{Introduction}
Distributed storage systems (DSS) have recently received significant research attention due to their important applications in data centers and cloud networks. To achieve reliability in DSS, some form of redundancy is introduced using either replication or erasure coding. Erasure coding is attractive in terms of storage efficiency, but it requires large amount of data to be downloaded during the repair of a failed node. To address this problem, Dimakis et al. \cite{Dimakis:07} introduced a new class of codes, referred to as {\it Regenerating Codes}, which significantly reduce the amount of data downloaded during the repair process. 

Specifically, Dimakis \etal~\cite{Dimakis:07}, \cite{Dimakis:10} showed that there exists a trade-off between storage space per node and repair bandwidth for single node failure, and that regenerating codes optimally achieve this trade-off. The codes on one extreme point of the optimal storage-repair bandwidth trade-off curve that minimize the repair bandwidth are referred to as Minimum Bandwidth Regenerating (MBR) codes; whereas, the codes on the other extreme point that minimize storage per node are referred to as Minimum Storage Regenerating (MSR) codes. Several explicit code constructions have been proposed for exact regenerating codes (see~\cite{Dimakis-Survey:11} and references therein). In this paper, we utilize product-matrix MBR codes~\cite{Rashmi:11}, since these codes can be constructed for the entire range of parameters and  require small finite field size.

Another important challenge for a DSS is the security of the stored data. For instance, some of the storage nodes in the cloud networks owned by certain private organizations can be eavesdropped. 
Providing secrecy against eavesdropping  is particularly challenging in DSS because of their dynamic nature, with nodes continually failing and being repaired. At any point of time, an eavesdropper {\it Eve} can observe any subset of nodes of bounded size. 

DSS can be secured using either conventional cryptographic techniques or information-theoretic approaches. One major drawback of almost all the secret key based encryption techniques is that they require secret key management mechanisms, which incur significant computational and communication overhead. Therefore, in the distributed setting of DSS, providing information-theoretic secrecy might be advantageous.

Information-theoretic model for securing regenerating codes was introduced by Pawar \etal~\cite{Pawar:10},~\cite{Pawar:11}. Since then, a number of investigations have been carried out on characterizing outer-bounds on secrecy capacity and the associated achievable schemes (see~\cite{Shah:11}, \cite{Zhu:13}, \cite{Rawat:14}). All these results are focussed on the paradigm of information-theoretic {\it perfect secrecy}. Essentially, perfect secrecy requires that the eavesdropper gains absolutely {\it no information} about the stored data from its observations. 

To be precise, suppose that a DSS is storing $\Bs$ message files $\Sf = \{S_1,\ldots,S_{\Bs}\}$, where each file can be considered as a symbol in a finite field $\GF{q}$. Let $\E$ denote the set of (encoded) symbols that Eve can observe. A DSS is said to be perfectly secure if the mutual information between the message symbols $\Sf$ and the eavesdropped symbols $\E$ is zero, {\it i.e.}, $\I{\Sf}{\E} = 0$. 

For many practical storage systems, this condition might be too strong. Moreover, coding schemes that provide perfect secrecy involve mixing data symbols with random keys to confuse the eavesdropper, which incurs loss in the storage capacity. Considering these drawbacks of the perfect secrecy notion, we focus on the notion of {\it weak secrecy}~\cite{Bhattad:05}.\footnote{Note that the notion of weak secrecy that is introduced in~\cite{Bhattad:05} and considered throughout this paper, is different from the conventional notion of information-theoretic weak secrecy, which is defined for asymptotically large block-lengths. The weak secrecy notion considered in this paper is applicable to finite block-lengths as well.} 

The notion of weak secrecy requires that an eavesdropper gains no information about any individual message file. For example, let the number of files be \mbox{$\Bs = 4$,} and the size of finite field be \mbox{$q=7$.} Further, suppose that Eve observes the following two encoded symbols \mbox{$\E  = \{S_1 + S_2 + S_3 + S_4,\: S_1 + 2S_2 + 3S_3 + 4S_4\}$.} Then, Eve cannot get any information about any individual message file, when the files are uniformly random and independent of each other. 

Furthermore, weak secrecy requires that even if Eve can obtain some $g$ number of files as a side information, it should not be able to decode for any other file. For instance, if Eve has a side-information of $g = 1$ file, she cannot decode for any other file observing $E$. Essentially, weakly secure coding schemes use data packets as keys, and thus, do not incur loss in capacity. 

Despite of its practical benefits, there have been relatively very few attempts on employing weak secrecy for DSS. In~\cite{Oliveira:12}, Oliveira \etal  ~present a construction of weakly secure erasure codes for DSS without considering the regeneration aspects. Very recently, Dau \etal~\cite{Dau:13} have analyzed the weak secrecy properties of two families of regenerating codes: regular-graph codes~\cite{Rashmi:09} and product-matrix codes~\cite{Rashmi:11}. 

In this paper, going a step ahead from~\cite{Dau:13}, we focus on designing outer codes to improve the weak secrecy properties of regenerating codes. To be specific, we present explicit construction of a coset coding based outer code to weakly secure product-matrix (PM) codes operating at MBR point (referred to as PM-MBR codes)~\cite{Rashmi:11} for the scenario wherein Eve can observe any single storage node. The proposed coding scheme has numerous advantages. First, it enhances the weak secrecy properties of the PM-MBR codes in terms of the amount of side-information $g$ that Eve can have without being able to decode any new file. In particular, when the size of the stored data is large, the gain in $g$ achieved by the proposed coding scheme is twofold. Second, the proposed outer codes leverage the elegant structure that is present in the PM codes, and thus, require small finite field size. Finally, the {\it weak-secrecy capacity} of the proposed coding scheme is {\it nearly equal}\footnote{For all parameters, the weakly secure capacity of the proposed scheme is two units below the non-secure storage capacity, which is negligible when the non-secure capacity is large.} to the non-secure storage capacity. These features make the proposed coding scheme attractive in practical settings.

\section{Preliminaries}
\label{sec:preliminaries}

\subsection{Regenerating Codes}
\label{sec:regen-codes}
Suppose we need to store a file $\Sf = \{S_1,\ldots,S_B\}$ containing $B$ symbols, each drawn uniformly and independently from a finite field $\GF{q}$, across $n$ storage nodes, where each node is capable of storing $\al$ symbols. A regenerating code encodes the $B$ message symbols into $n\alpha$ symbols in such a way that it satisfies the following two properties.
First, a {\it data collector} (DC) connecting to \emph{any} $k$ out of $n$ nodes is able to reconstruct the entire file; this is referred to as  the {\it reconstruction property}. Second, when a storage node is failed, it is \emph{regenerated} by adding a new node which downloads $\beta$ symbols each from any $d$ out of the remaining $n - 1$ nodes; this is referred to as the {\it regeneration property}. 
A regenerating code with these parameters is referred to as an $(n,k,d,\al,\be)$ regenerating code. 

Under these requirements, the outer bound on the capacity of an $(n,k,d,\al,\be)$ regenerating code is given as~\cite{Dimakis:10}
\begin{equation}
\label{eq:capacity}
B \leq \sum_{i=0}^{k-1}\min\{\alpha,(d-i)\beta\}
\end{equation}
For Minimum Bandwidth Regenerating (MBR) codes, first the {\it repair bandwidth} $d\beta$ is minimized and then the storage per node $\alpha$ is minimized.
Specifically, for an $(n,k,d,\al,\be = 1)$ MBR code, we have 
$B = \sum_{i=0}^{k-1} (d-i)$, and $\al = d$.
If the regenerated node is an exact replica of the failed node, then the repair model is said to be {\it exact repair}~\cite{Rashmi:09}. In this paper, we focus on a special class of exact minimum bandwidth regenerating (MBR) codes called as the product-matrix codes~\cite{Rashmi:11}, which are described in section~\ref{sec:PM-codes}.

\subsection{Eavesdropper Model}
\label{sec:Eve-model}
The most generalized eavesdropper model for a DSS, called as the $(l_1,l_2)$-eavesdropper model, is considered in \cite{Rawat:14} (see also~\cite{Shah:11}). An $(l_1,l_2)$-eavesdropper, {\it Eve}, can access the data stored on any $l_1$ nodes, and the data downloaded during the regeneration of any $l_2$ nodes. 

Notice that at MBR point, the number of downloaded symbols is equal to the number of stored symbols. Therefore, Eve cannot gain any additional information by observing the data downloaded during the regeneration, and thus, it is sufficient to simply consider the total number of nodes $l := l_1 + l_2$ that Eve has access to. 

In this paper, we assume that Eve can access any one storage node. Thus, we have $l = 1$.
We assume that Eve is passive, has unbounded computational power, and has the knowledge of the coding scheme being used.

\subsection{Information-theoretic Secrecy}
\label{sec:secrecy}
Suppose $\Sf = \{S_1,\ldots,S_{\Bs}\}$ denote the $\Bs$ message files where each file $S_i\in\GF{q}$, and $\E$ denotes Eve's observations. A DSS is said to be {\it perfectly secure} if $\I{\Sf}{\E} = 0$. Under this requirement, Pawar {\it et al.}~\cite{Pawar:10} characterized an upper bound on the {\it secrecy capacity} as:
\begin{equation}
\label{eq:capacity-secure}
\Bs \leq \sum_{i=l}^{k-1}\min\{\alpha,(d-i)\beta\}
\end{equation}
Comparing \eqref{eq:capacity} and \eqref{eq:capacity-secure}, we can say that in a perfectly secure DSS, the $l$ nodes that are compromised by the eavesdropper cannot effectively contain any useful information. Consequently, the perfect secrecy requirement results in a loss of storage capacity, {\it i.e.}, $\Bs < B$. 
\begin{remark}
\label{rem:secure-PM-MBR}
Shah \etal~\cite{Shah:11} show that PM-MBR codes can be made perfectly secure against an $(l_1,l_2)$ eavesdropper by appropriately mixing random keys with the message symbols. The secure codes achieve the capacity outer bound given in~\eqref{eq:capacity-secure}, and the loss of capacity incurred due to perfect secrecy requirement is $B-\Bs = ld - \binom{l}{2}$.
\end{remark}

In this paper, we focus on a relaxed, yet practically appealing, notion of {\it weak secrecy}~\cite{Bhattad:05}. A DSS is said to be {\it weakly secure} if $\I{S_i}{\E} = 0$, $\forall i\in[\Bs]$, where $[\Bs] := \{1,\ldots,\Bs\}$. Furthermore, suppose Eve is able to obtain, as a side information, some $g$ message symbols denoted as $\SG := \{S_i : i\in\G\}$ for some $\G\subset[\Bs]$ such that $\left|\G\right| = g$. Then, a DSS is said to be {\it weakly secure against $g$ guesses} if we have
\begin{equation}
\label{eq:g-secure}
\I{S_i}{\E | \SG}  =  0 \qquad \forall i\in[\Bs]\setminus\G,\: \forall\G\subset[\Bs] : \left|\G\right| \leq g.
\end{equation}
In \cite{Silva:11}, it was shown that the above condition is equivalent to
\begin{equation}
\label{eq:g1-block-secure}
\I{\SGprime}{\E}  =  0 \qquad \forall{\Gprime}\subseteq[\Bs] : \left|\Gprime\right| \leq g+1.
\end{equation}
Essentially, this condition implies that in a scheme that is weakly secure against $g$ guesses, it is not possible for Eve to obtain any information about {\it any} subset of $g+1$ symbols from her observations.\footnote{In~\cite{Dau:13}, a scheme that is weakly secure against $g-1$ guesses is called as a $g$-{\it block secure} scheme, following condition~\eqref{eq:g1-block-secure} as a definition of $g$-block security.} 

\subsection{Recap of Product-Matrix MBR Codes}
\label{sec:PM-codes}
Let us review the Product-Matrix framework based MBR Codes (PM-MBR Codes) proposed in \cite{Rashmi:11}. 
The PM-MBR code is obtained by taking the product of an $(n\times d)$ encoding matrix $\Psi$ and a $(d\times\alpha)$ message matrix $M$ that contains the $B$ message symbols arranged in a particular fashion. Specifically, the encoding matrix $\Psi$ and the message matrix $M$ have the following structure
\begin{equation}
\label{eq:PM-code}
\underbrace{\Psi}_{n\times d} =
\begin{bmatrix}
\underbrace{\Phi}_{n\times k} & \underbrace{\Delta}_{n\times (d-k)}
\end{bmatrix},\:\:
\underbrace{M}_{d\times d} =
\begin{bmatrix}
\underbrace{M_1}_{k\times k} & \underbrace{M_2}_{k\times (d-k)} \\
\underbrace{M_2^T}_{(d-k)\times k} & \underbrace{0}_{(d-k)\times (d-k)}
\end{bmatrix}
\end{equation}

In the message matrix $M$, the component matrix $M_1$ is a $k\times k$ symmetric matrix which contains $\frac{k(k+1)}{2}$ data symbols in the upper triangular half; whereas, the other component matrix $M_2$ is a $k\times (d-k)$ matrix which contains the remaining $k(d-k)$ message symbols. Note that at MBR point, $B = \sum_{i=0}^{k-1}(d-i) = \frac{k(k+1)}{2}+k(d-k)$. 

The matrices $\Phi$ and $\Delta$ are chosen in such a way that any $k$ rows of $\Phi$ are linearly independent, and any $d$ rows of $\Psi$ are linearly independent. If $\Psi$ is chosen to be a Vandermonde or a Cauchy matrix, these requirements are satisfied. Note that the field size $q$ depends on the choice of the encoding matrix $\Psi$. For instance, if $\Psi$ is a Vandermonde matrix, then field sixe of $q\geq n$ is both necessary and sufficient. 

The $\alpha$ symbols stored on the $i$-th node are given by $C_i = \Psi_i M$, where $\Psi_i$ denotes the $i$-th row of $\Psi$. The regeneration and the reconstruction processes can be found in \cite{Rashmi:11}. 

{\it Example:} Consider a $(n=5,k=3,d=4,\al=4,\be=1)$ PM-MBR code. Then, from~\eqref{eq:capacity}, we have $B = 9$. Let the data to be stored is given as $\X = \{x_1,\cdots,x_9\}$, where $x_i\in\GF{q}$ $\forall i$. Suppose the encoding matrix $\Psi$ is a Cauchy matrix. Then, in parametric form, we have
\begin{equation}
\label{eq:MBR-ex}
\Psi = 
\left[\frac{1}{a_i+b_j}\right]_{i=1, j=1}^{5,4},\:\:
M = 
\begin{bmatrix}
x_1 & x_2 & x_3 & x_4\\
x_2 & x_5 & x_6 & x_7\\
x_3 & x_6 & x_8 & x_9\\
x_4 & x_7 & x_9 & 0
\end{bmatrix},
\end{equation}
where $a_i, b_j\in\GF{q}$ such that $a_i\neq b_j$ and $a_i + b_j \neq 0$ for all $i, j$. Note that, to satisfy these requirements, we need at least $n+d=9$ distinct elements, and thus, we require $q\geq 9$.

\section{Explicit Code Construction for Weak Security}
\label{sec:main}

\subsection{Summary of Main Results}
\label{sec:summary}
We propose an explicit construction of coset coding based outer code for PM-MBR inner code that achieves weak secrecy when Eve can observe any single storage node, \ie, $l = 1$. The proposed scheme works for the entire range of parameters $[n, k, d]$ that are feasible for DSS. The weak secrecy capacity of the proposed scheme is $\Bs = B - 2$, where $B$ is the capacity without any secrecy requirement. The proposed scheme is weakly secure against $g \leq d+k-4$ number of guesses.

\subsection{Outer Code Based on Coset Coding}
\label{sec:PM-secure}
We propose to use an outer code to improve the weak secrecy level (\ie, the amount of side information that Eve can have) of the PM-MBR codes. When outer code is used, the overall encoding consists of two steps. First, an outer code is used to encode the length-$\Bs$ message file $\Sf=\left[S_1\ldots S_{\Bs}\right]\in\GF{q}^{\Bs}$ into a codeword $\X=\left[X_1\ldots X_B\right]\in\GF{q}^{B}$. Next, the codeword $X$ is encoded using the PM-MBR code (as an inner code) by populating the entries of matrix $M$ (see~\eqref{eq:PM-code}) with codeword symbols $X$. Notice that the regeneration process remains the same. To obtain the message file $\Sf$, a user would first decode $\X$ using the reconstruction process of the PM-MBR code, and then, decode the outer code to get $\Sf$.

We design the outer code based on {\it coset coding}~\cite{Ozarow:84}. 
A coset code is constructed using a $(B,B-\Bs)$ linear code $\code$ over $\GF{q}$ with parity-check matrix $H\in\GF{q}^{\Bs\times B}$. Specifically, the message file $\Sf$ is encoded by selecting uniformly at random some $\X\in\GF{q}^B$ such that $\Sf = H\X$. Therefore, the message file can be considered as a syndrome specifying a coset of $\code$, and the codeword $\X$ is a randomly chosen element of that coset. Notice that the decoding operation of a coset code consists of simply computing the syndrome $\Sf = H\X$.

To design the matrix $H$ appropriately, we need to transform the weak secrecy condition~\eqref{eq:g1-block-secure} into a condition involving $H$. For this, we use the following result from~\cite[Lemma 6]{Silva:11}, which is a generalization of~\cite[Theorem 1]{Salim:07}.
\begin{lemma}
\label{lem:silva}
(\cite{Silva:11}) Suppose a coset code based on a parity-check matrix $H\in\GF{q}^{\Bs\times B}$ is used as an outer code over a given exact regenerating code to store the message $\Sf = \{S_1,\cdots, S_{\Bs}\}$. Suppose each message symbol $S_i$ for $i \in [\Bs]$ is chosen uniformly and independently. Let $\E = G\X$ be the $\mu$ linearly independent symbols observed by an eavesdropper. Then, for any $\Gprime\subseteq[\Bs] : \left|\Gprime\right| \leq B - \mu$, we have
\begin{equation}
\label{eq:subspace-condition-0}
\I{\SGprime}{\E} = \rank\: H_{\Gprime} + \rank\: G - \rank\:\twomatrix{H_{\Gprime}}{G},
\end{equation}
where $H_{\Gprime}$ is a sub-matrix of $H$ formed by choosing the rows indexed by the set $\Gprime$.
\end{lemma}
Then, weak secrecy would be satisfied by designing $H$ and $G$ such that 
\begin{equation}
\label{eq:subspace-condition}
\rank\:\twomatrix{H_{\Gprime}}{G} = \rank\: H_{\Gprime} + \rank\: G, \:\: \forall \Gprime\subset[\Bs] : \left|\Gprime\right| \leq g+1.
\end{equation}

\subsection{Outer Code Construction for PM-MBR Codes}
\label{sec:securing-PM_MBR}
As previously mentioned in section~\ref{sec:Eve-model}, we assume that Eve can observe any single storage node.
Let $e$ denote the index of the node that Eve can access. Eve observes $\alpha=d$ symbols stored on node $e$ given by $E = \psi_e M$, where $\psi_e$ is the $e$-th row of $\Psi$. To use condition~\eqref{eq:subspace-condition}, we need to find a matrix $G_e$ such that $E = G_eX$. 
This is carried out by a simple linear transformation that guarantees $E = (\psi_e M)^T = G_e\X$. 

To describe the transformation formally, assume without loss of generality, that the $B$ outer-coded symbols $\X= \{\X_1,\ldots,\X_B\}$ are filled in the message matrix $M$ in a lexicographic order for $1\leq j\leq d$ and $1\leq i\leq k$. Therefore, if $M_{(i,j)}$ denotes the symbol at $i$-th row and $j$-th column of $M$, then we have $M_{(1,1)} = \X_1$, $M_{(1,2)} = \X_2$, $\ldots\:$, $M_{(k,d)} = \X_B$. Equivalently, we have $\X = \{M_{(1,1)},\ldots, M_{(1,d)}, M_{(2,2)}, \ldots, M_{(2,d)}, \ldots, M_{(k,k)}, \ldots, M_{(k,d)}\}$. Further, notice that $M$ is a symmetric matrix. Thus, if $M_{(i,j)} = \X_b$, then we have $M_{(j,i)} = \X_b$ as well. 

Under this setting, the symbols observed by Eve can be written as $E = G_e X$,
where $(i,b)$-th entry of the $d\times B$ matrix $G_{e}$ is given as
\begin{equation}
\label{eq:G-e}
G_{e}(i,b) = 
\begin{cases}
\Psi_{(e,j)} & \textrm{if} \: M_{(i,j)} = \X_b \: \textrm{for some}\: j \in[d],\\
0 & \textrm{otherwise},
\end{cases}
\end{equation}
for $1\leq i\leq d$ and $1\leq b\leq B$. 
Note that $G_e$ can be considered as a generator matrix of PM-MBR code for node $e$.
To ensure weak security against $g$ guesses, we need to design $H$ such that for each node $e\in[n]$, its generator matrix $G_e$ satisfies~\eqref{eq:subspace-condition}. 

{\it Example:} For the previous example of $(n=6,k=4,d=5,\al=5,\be=1)$ PM-MBR code, if Eve observes the first node then we can write $G_{1}$ as~\eqref{eq:G-E-ex} (shown at the top of the page).
\begin{equation}
\label{eq:G-E-ex}
G_{1} = 
\begin{bmatrix}
\Psi(1,1) & \Psi(1,2) & \Psi(1,3) & \Psi(1,4) & 0 & 0 & 0 & 0 & 0\\
0 & \Psi(1,1) & 0 & 0 & \Psi(1,2) & \Psi(1,3) & \Psi(1,4) & 0 & 0\\
0 & 0 & \Psi(1,1) & 0 & 0 & \Psi(1,2) & 0 & \Psi(1,3) & \Psi(1,4)\\
0 & 0 & 0 & \Psi(1,1) & 0 & 0 & \Psi(1,2) & 0 & \Psi(1,3)
\end{bmatrix}
\end{equation}

\begin{remark}
\label{rem:weak-security-PM-MBR}
Observe that the matrix $G_e$ for each node $e\in[n]$ is sparse. In particular, $G_e$ for each node $e\in[n]$ contains at least one row vector with Hamming weight $k$. Thus, PM-MBR codes are not secure against $g\geq k-1$ guesses, when Eve can observe one storage node. This shows the necessity to employ an outer code to improve the level of weak secrecy.
\end{remark}

\begin{remark}
\label{rem:random-H}
It is possible to use a random matrix as $H$, however it would require very large field size. This is because the condition~\eqref{eq:subspace-condition} must be satisfied for all sub-matrices of $H$, the number of which are exponentially large. Moreover, for each sub-matrix $H_{\Gprime}$, we must ensure~\eqref{eq:subspace-condition} for each node $e\in[n]$, since Eve can observe any storage node. Therefore, 
we explicitly construct $H$ that 
requires small field size.
\end{remark}

Our aim is to jointly design a PM-MBR code and a coset code such that~\eqref{eq:subspace-condition} is satisfied. Notice that while designing PM-MBR codes, the only degree of freedom that we have is in choosing the encoding matrix $\Psi$ such that the conditions specified in section~\ref{sec:PM-codes} are satisfied. 

The main idea of our solution is to construct $H$ such that it has the same {\it structure} as that of the generator matrix $G_{e}$ of a node for the PM-MBR code. The same structure of $G_e$ and $H$ would enable us to ensure the condition~\eqref{eq:subspace-condition}. 

Since $\G_e\X = \psi_e M$, the values of the non-zero entries in $G_e$ are specified by $\psi_e$ and their locations depend on the elements of $M$. Further, the location of non-zero entries in $G_e$ are the same for all nodes $e\in[n]$. To formally specify this {\it structure} present in $G_e$, we introduce the notion of {\it type}. We say that a length-$B$ row vector $h^{(j)}$ is of type $j$ if the indices of its non-zero coefficients are the same as that of $i$-th row of $G_{e}$; the values of the non-zero coefficients can be different. We call the corresponding set of indices of non-zero coefficients as the {\it index set of type} $j$, denoted as $\Ind{j}$. Observe that, essentially, the type of a vector specifies the locations of the non-zero coefficients of the vector. Further, the total number of types is equal to the number of rows of $G_e$ which is $d$.

{\it Example:} Considering our running example, a vector of type 4 has the form $h^{(4)} = \begin{bmatrix}0 & 0 & 0 & \gamma & 0 & 0 & \gamma^2 & 0 & \gamma^3\end{bmatrix}$ for some $\gamma\in\GF{q}$. The corresponding index set of type 4 is $\Ind{4} = \{4, 7, 9\}$, which corresponds to the indices of elements of fourth column of $M$ (see~\eqref{eq:MBR-ex}). 

We construct $H$ such that each row of $H$ belongs to one of the $d$ types.
Let $\theta_i$ denote the number of row vectors of type $i$, $1\leq i\leq d$, that are present in $H$. Define $\theta := [\theta_1 \cdots \theta_d]$, which we call as the {\it type cardinality vector}. For each $\theta_i >0$, let $H_{i}$ denote the $\theta_i \times B$ sub-matrix of $H$ that is composed of all row vectors of type $i$.

Once the type of a row vector is fixed, it is sufficient to give a set of values of non-zero coefficients to fully specify the row vector. 
For example, the non-zero coefficient values of all the vectors in $G_e$ are specified by the row vector $\psi_e$. In a similar manner,  we represent the non-zero coefficients of all the row vectors in $H$ using a matrix $\Psih$. Specifically, a $d\times d$ matrix $\Psih$ is defined in such a way that the $j$-th row of $\Psih$ specifies the non-zero coefficient values of the $j$-th vector of type $i$ that is present in $H$ for $i, j\in[d]$. We call the matrix $\Psih$ as the {\it coefficient matrix}.
Observe that the type cardinality vector $\theta$ along with the coefficient matrix $\Psih$ are sufficient to specify the parity-check matrix $H$. 

In the following, we describe an explicit construction of the encoding matrix $\Psi$ and the parity check matrix $H$ (in terms of $\theta$ and $\Psih$), which improves the security performance of the PM-MBR codes beyond $g = k-1$ guesses (see remark~\ref{rem:weak-security-PM-MBR}). 

\begin{construction}
\label{con:H-PM-MBR}
First, choose the type cardinality vector $\theta$ as follows.
\begin{equation}
\label{eq:card-theta-PM-MBR}
\theta_i =
\begin{cases}
0 & \textrm{if} \quad i = 1,\\
d-k+i & \textrm{if} \quad 2\leq i\leq k-1,\\
d-1 & \textrm{if} \quad i = k,\\
1 & \textrm{if} \quad k+1\leq i\leq d.
\end{cases}
\end{equation}
Note that $\max_{1\leq i\leq d} \theta_i = d-1$. 

Next, choose an $n\times d$ encoding matrix $\Psi$ and a $d\times d$ coefficient matrix $\Psih$ in such a way that any square sub-matrix of $\Psit := \twomatrix{\Psi}{\Psih}$ is non-singular. 

Finally, using $\Psih$ and $\theta$, construct $H$ as follows. For each $\theta_i$, $2\leq i\leq d$, the $\theta_i \times B$ sub-matrix $H_i$ of $H$ is given as  
\begin{equation}
\label{eq:H-from-theta}
H_i (p,b) = 
\begin{cases}
\Psih_{(p,j)} & \textrm{if} \: M_{(i,j)} = \X_b \: \textrm{for some}\: j \in[d],\\
0 & \textrm{otherwise},
\end{cases}
\end{equation}
for $1\leq p\leq \theta_i$ and $1\leq b\leq B$. The parity-check matrix $H$ is obtained by vertically concatenating the sub-matrices $H_i$, {\it i.e.}, $H = \left[H_2^T \: H_3^T \: \ldots \: H_d^T\right]^T$.
\end{construction}

Note that the requirement on $\Psit$ mentioned in Construction~\ref{con:H-PM-MBR}, that any of its square sub-matrices should be non-singular, can be ensured, for example, by choosing $\Psit$ as a Cauchy matrix. Another construction of a matrix that satisfies this requirement can be found in~\cite{Lacan:04}. Both these constructions require that $q\geq n + 2d$.

There are couple of points about this requirement on $\Psit$ that are worth mentioning. First, note that this requirement on $\Psit$ implies that any square sub-matrix of $\Psi$ should also be non-singular. This is a stronger requirement, which guarantees the two requirements on $\Psi$ that are mentioned in section~\ref{sec:PM-codes}. Second, choosing $\Psi$ as a Vandermonde matrix, which is good enough to meet the requirements of PM-MBR codes, is not sufficient, since a Vandermonde matrix defined over finite field can contain singular square sub-matrices (see~\cite{Lacan:04}, also~\cite{Dau:13}).  
 
{\it Example:} For the running example, construction~\ref{con:H-PM-MBR} yields $\theta = \{ 0, 3, 3, 1\}$. Let the $5\times 4$ encoding matrix $\Psi$ be a Cauchy matrix ({\it cf.}~\eqref{eq:MBR-ex}). We choose $\Psih$ such that $\Psih = \twomatrix{\Psi}{\Psih}$ is also a Cauchy matrix. Note that this requires $q\geq 13$. Then, the resulting parity-check matrix $H$ is given in~\eqref{eq:H-ex}.

\begin{equation}
\label{eq:H-ex}
H = 
\begin{bmatrix}
0 & \Psih(1,1) & 0 & 0 & \Psih(1,2) & \Psih(1,3) & \Psih(1,4) & 0 & 0\\
0 & \Psih(2,1) & 0 & 0 & \Psih(2,2) & \Psih(2,3) & \Psih(2,4) & 0 & 0\\
0 & \Psih(3,1) & 0 & 0 & \Psih(3,2) & \Psih(3,3) & \Psih(3,4) & 0 & 0\\
0 & 0 & \Psih(1,1) & 0 & 0 & \Psih(1,2) & 0 & \Psih(1,3) & \Psih(1,4)\\
0 & 0 & \Psih(2,1) & 0 & 0 & \Psih(2,2) & 0 & \Psih(2,3) & \Psih(2,4)\\
0 & 0 & \Psih(3,1) & 0 & 0 & \Psih(3,2) & 0 & \Psih(3,3) & \Psih(3,4)\\
0 & 0 & 0 & \Psih(1,1) & 0 & 0 & \Psih(1,2) & 0 & \Psih(1,3)
\end{bmatrix}
\end{equation}

\subsection{Analysis}
\label{sec:analysis}
First, we characterize the file size that can be stored in a weakly secure sense by using the proposed outer code along with a PM-MBR code.
\begin{theorem}
\label{thm:secure-capacity}
When an outer coset code based on the parity-check matrix $H$ given in Construction~\ref{con:H-PM-MBR} is used along with a PM-MBR code, the weakly secure storage capacity is \mbox{$\Bs = B - 2$.}
\end{theorem}
\begin{IEEEproof}
See appendix   
\end{IEEEproof} 


Next, we compute the level of secrecy that can be attained using the proposed outer code along with a PM-MBR code.
\begin{theorem}
\label{thm:guesses-allowed}
An outer coset code based on the parity-check matrix $H$ given in Construction~\ref{con:H-PM-MBR} makes a PM-MBR code weakly secure against $g \leq d + k - 4$ guesses.
\end{theorem}
\begin{IEEEproof}
See Appendix
\end{IEEEproof}

\begin{remark}
\label{rem:comparison-Dau}
In~\cite{Dau:13}, it is shown that, when Eve observes $l$ nodes, the PM-MBR codes using Cauchy matrix as their encoding matrix are weakly secure against $k-l-1$ guesses. Thus, for $l=1$, it is shown that PM-MBR codes are secure against $k-2$ guesses. Our proposed encoding enhances the level of security to $d+k-4$ guesses, which is an improvement of $d-2$ for all set of parameters (except for $d=k=2$). Notice that, for any regenerating code, $d\geq k$. Thus, for large $k$, the enhancement achieved by the proposed scheme is almost twofold in terms of the number of guesses that Eve can make and still cannot learn anything about any single message symbol.
\end{remark}

\appendices
\section{Proof of Theorem~\ref{thm:secure-capacity}}
\label{app:proof-1}

Notice that the message file, which is securely stored, can be considered as the syndrome of the coset code as $S=HX$. Thus, the weak-secrecy capacity is the dimension of matrix $H$. First, we show that, if $H$ is designed following construction~\ref{con:H-PM-MBR}, it contains $B-2$ rows. Next, we show that $H$ is full-rank to prove the result.

Now, notice that the total number of rows in $H$ is equal to $\sum_{i=1}^d \theta_i$. From~\eqref{eq:card-theta-PM-MBR}, we have
\begin{IEEEeqnarray}{rCl}
\sum_{i=1}^d \theta_i & = & 0 + \left(\sum_{i=2}^{k-1}d-k+i\right) + (d-1) + (d-k) \nonumber\\
& = & \left(\sum_{i=1}^{k-2} d-i \right)+ (d-1) + (d-k) \nonumber\\
& \stackrel{(a)}{=} & \left(\sum_{i=0}^{k-1}d-i\right) -2 \nonumber\\
& \stackrel{(b)}{=} & B - 2
\end{IEEEeqnarray} 
where (a) can be easily obtained by carrying out simple algebraic manipulations, and (b) follows from~\eqref{eq:capacity} and from the fact that at MBR point $\alpha = d$ (we assume that $\be = 1$). 

To prove that $H$ is full-rank, we show that it is possible to append two rows to $H$ in such a way that the resulting $B\times B$ matrix, denoted as $H'$, is non-singular. Specifically, append a type 1 row vector with non-zero coefficients corresponding to the first row of $\Psih$, and append a type $k$ row vector with non-zero coefficients corresponding to the $d$-th row of $\Psih$. From~\eqref{eq:card-theta-PM-MBR}, it is easy to see that $H'$ contains $d - (k  - i)$ rows of type $i$ for $k\geq i\geq 2$, one row of type 1, and one row each of types $k+1$ through $d$. Without loss of generality, assume that the rows of $H$ are ordered in such a way that first $d$ rows are of type $k$, next $d-1$ rows are of type $k-1$ and so on up to $d-(k-2)$ rows of type 2. The last $d-k+1$ rows are of types $k+1$ through $d$ and of type 1, respectively.

Now, for proving non-singularity of $H'$, consider a system of linear equations $Z = H'Y$, where $Y = [Y_1 \cdots Y_B]$ and $Z=[Z_1 \cdots Z_B]$ are length-$B$ vectors of variables $Y_1$ through $Y_B$ and $Z_1$ through $Z_B$, respectively. We show that it is possible to {\it successively decode} variables in $Y$ in terms of variables in $Z$ by leveraging the elegant structure of $H'$. 

To describe the process of successive decoding, we need to introduce some notation.
Recall that the type of a row vector specifies the locations of the non-zero coefficients of the vector.  The corresponding set of locations of non-zero coefficients of a type $i$ vector is referred to as the index set of type $i$, denoted as $\Ind{i}$. Define $Y[\Ind{i}] := \{Y_l : l\in\Ind{i}\}$, {\it i.e.}, $Y[\Ind{i}]$ is the vector of elements of $Y$ that are indexed by the index set of type $i$. 

Suppose we write vector $Y$ in the format of matrix $M$ (see~\eqref{eq:PM-code}) in a lexicographic order.
Notice that the index set of type $i$ is the set of indices of the elements in the $i$-th column of $M$. 
Observing the structure of $M$, we divide the $d$ types into two groups. The types $1$ through $k$ are called as group I, while the types $k+1$ through $d$ are called as group II. 
For any group I type, the index set consists of $d$ elements, \ie, $|\Ind{i}| = d$, $\forall i\in[k]$. On the other hand, for any group II type, the corresponding index set has $k$ elements, \ie, $|\Ind{i}| = k$ $\forall i\in\{k+1,\ldots,d\}$. 
Further, index set corresponding to any group I has one index common with the index sets of all other types, \ie, $|\Ind{i}\cap\Ind{j}| = 1$ $\forall i< j: i\in[k]$. Whereas, any pair of index sets of group II types has no common symbol, \ie, $|\Ind{i}\cap\Ind{j}| = 0$ $\forall k< i < j\leq d$.

Let $\gam_1$ and $\gam_2$ denote the number of row vectors in $H'$ of group I and group II types, respectively. 
Algorithm~\ref{alg:succ-decoding} presented below decodes elements of $Y[\Ind{i}]$ for each $i\in [d]$ successively.

\begin{algorithm}[!t]
\caption{Successive decoding for $Z=H'Y$}
\label{alg:succ-decoding}
\begin{algorithmic}[1]
\STATE Set $\gam_1 = d$, $\gam_2 = k$
\FOR{$p = k$ \TO 2}
\STATE{Consider set of equations corresponding to $\gam_1$ rows of type $p$ as $Z[\Ind{p}] = \Psih_{1:\gam_1}Y[\Ind{p}]$}
\STATE{Using perviously decoded elements, decode for elements of $Y$ located at \mbox{$\Ind{p}\setminus\left(\bigcup_{l=1}^{k-p}\left(\Ind{p}\cap\Ind{p+l}\right)\right)$}}
\STATE{$\gam_1 = \gam_1 - 1$, $\gam_2 = \gam_2 - 1$}
\ENDFOR 
\STATE{Decode for the remaining elements in index sets of types $k+1$ through $d$}
\STATE{Decode for the remaining single element of type $1$}  
\end{algorithmic}
\end{algorithm}

\begin{claim}
\label{clm:correctness-1}
Algorithm~\ref{alg:succ-decoding} decodes all the $B$ elements of $Y$ in terms of $Z$.
\end{claim}
\begin{IEEEproof}
The algorithm begins with type $k$, of which there are $d$ row vectors in $H'$. By the construction of $H'$, the non-zero coefficient values are the elements of the $d\times d$ Cauchy matrix $\Psih$. Thus, it is possible to solve for $Y[\Ind{k}]$ by inverting $\Psih$. Next, we prove by induction that, for $2\leq i\leq k$, if the elements of $\Ind{i+1}$ through $\Ind{k}$ have been decoded, then it is possible to decode the elements of $\Ind{i}$. By construction of $H'$, there are $d-k+i$ rows of type $i$ in $H'$ for $2\leq i\leq k$ with non-zero coefficients given by $\Psih_{1:(d-k+i)}$, respectively. This forms a system of $d-k+i$ linear equations in $d$ variables of $\Ind{i}$ as $Z[\Ind{i}]=\Psih_{1:(d-k+i)}Y[\Ind{i}]$. Note that, since type $i$ is a group I type, there is one element common between $\Ind{i}$ and each of $\Ind{i+1}$ through $\Ind{k}$. Thus, there are $k-i$ elements in $\Ind{i}$ that have already been decoded. As any square sub-matrix of $\Psih$ is non-singular by construction (it is a Cauchy matrix), it is possible to solve for the un-decoded $d-k+i$ variables using $Z[\Ind{i}]=\Psih_{1:(d-k+i)}Y[\Ind{i}]$. 

At the end of the first \textbf{for} loop, $k-1$ elements from $\Ind{j}$ for each $j\in[d]$ are decoded. Thus, in each of the index sets of group II types, there is just one element to be decoded. By construction, $H'$ has one row in each of the group II types, and thus, it is possible to decode all the elements in group II index sets. Note that, at this point, all the elements from index sets of all types except type 1 are decoded.

Finally, since $\Ind{1}$ has one element common with all the remaining $d-1$ index sets, only single element from $\Ind{1}$ remains to be decoded, which can be decoded using a row of type 1 that is appended to $H$.

Notice that, as each index set corresponds to a column of matrix $M$, we have $\bigcup_{j=1}^d \Ind{j}= \{Y_1, \ldots, Y_B\}$. Therefore, algorithm~\ref{alg:succ-decoding} decodes all the $B$ elements of $Y$. 
\end{IEEEproof}

\begin{remark}
\label{rem:alg1}
Note that successively decoding for the variables of a particular type is equivalent to performing Gaussian elimination on the corresponding rows of that particular type. Thus, in essence, the procedure for successive decoding gives the order in which Gaussian elimination can be performed in $H'$.
\end{remark}

\emph{Example:} Consider the following $H$ given in~\eqref{eq:H-ex-1}, which follows from construction~\ref{con:H-PM-MBR}. To form $H'$, first append a row vector of type 3 with non-zero coefficients specified by $\Psih_4$. Then, append a row vector of type 1 with non-zero coefficients specified by $\Psih_1$. See equation~\eqref{eq:Hp-ex-1}. To prove that $H'$ is full-rank, observe that we can decode for variables indexed by $\Ind{3} = \{2, 5, 6, 7\}$ using the four rows of type 3. Notice that $\Ind{3}\cap\Ind{2} = 6$. Then, using the three rows of type 2 and already decoded variable at index 6, solve for variables indexed by $\Ind{2}\setminus(\Ind{3}\cap\Ind{2}) = \{3, 8, 9\}$. Then, using the row vector of type 4, decode for $\Ind{4}\setminus((\Ind{4}\cap\Ind{3})\cup(\Ind{4}\cap\Ind{2})) = \{4\}$. Finally, using the row of type one, decode for $\Ind{1}\setminus((\Ind{1}\cap\Ind{4})\cup(\Ind{1}\cap\Ind{3})\cup(\Ind{1}\cap\Ind{2})) = \{1\}$. The successive decoding uses the property that any square sub-matrix of the Cauchy matrix $\Psih$ is non-singular.
\begin{equation}
\label{eq:H-ex-1}
H = 
\begin{bmatrix}
0 & \Psih(1,1) & 0 & 0 & \Psih(1,2) & \Psih(1,3) & \Psih(1,4) & 0 & 0\\
0 & \Psih(2,1) & 0 & 0 & \Psih(2,2) & \Psih(2,3) & \Psih(2,4) & 0 & 0\\
0 & \Psih(3,1) & 0 & 0 & \Psih(3,2) & \Psih(3,3) & \Psih(3,4) & 0 & 0\\
0 & 0 & \Psih(1,1) & 0 & 0 & \Psih(1,2) & 0 & \Psih(1,3) & \Psih(1,4)\\
0 & 0 & \Psih(2,1) & 0 & 0 & \Psih(2,2) & 0 & \Psih(2,3) & \Psih(2,4)\\
0 & 0 & \Psih(3,1) & 0 & 0 & \Psih(3,2) & 0 & \Psih(3,3) & \Psih(3,4)\\
0 & 0 & 0 & \Psih(1,1) & 0 & 0 & \Psih(1,2) & 0 & \Psih(1,3)
\end{bmatrix}
\end{equation}
\begin{equation}
\label{eq:Hp-ex-1}
H' = 
\begin{bmatrix}
0 & \Psih(1,1) & 0 & 0 & \Psih(1,2) & \Psih(1,3) & \Psih(1,4) & 0 & 0\\
0 & \Psih(2,1) & 0 & 0 & \Psih(2,2) & \Psih(2,3) & \Psih(2,4) & 0 & 0\\
0 & \Psih(3,1) & 0 & 0 & \Psih(3,2) & \Psih(3,3) & \Psih(3,4) & 0 & 0\\
0 & 0 & \Psih(1,1) & 0 & 0 & \Psih(1,2) & 0 & \Psih(1,3) & \Psih(1,4)\\
0 & 0 & \Psih(2,1) & 0 & 0 & \Psih(2,2) & 0 & \Psih(2,3) & \Psih(2,4)\\
0 & 0 & \Psih(3,1) & 0 & 0 & \Psih(3,2) & 0 & \Psih(3,3) & \Psih(3,4)\\
0 & 0 & \Psih(4,1) & 0 & 0 & \Psih(4,2) & 0 & \Psih(4,3) & \Psih(4,4)\\
0 & 0 & 0 & \Psih(1,1) & 0 & 0 & \Psih(1,2) & 0 & \Psih(1,3)\\
\Psih(1,1) & \Psih(1,2) & \Psih(1,3) & \Psih(1,4) & 0 & 0 & 0 & 0 & 0
\end{bmatrix}
\end{equation}


\section{Proof of Theorem~\ref{thm:guesses-allowed}}
\label{app:proof-2}

Essentially, we need to prove that condition~\eqref{eq:subspace-condition} always holds for the proposed coding scheme as long as $|\Gprime| \leq d + k - 3$. For notational convenience, let $T := \twomatrix{H_{\Gprime}}{G_e}$. Notice that there are $\binom{\Bs}{|\Gprime|}$ number of ways to choose a particular $|\Gprime|$, and we ensure~\eqref{eq:subspace-condition} for each possible $H_{\Gprime}$ as long as $|\Gprime| \leq d+k-3$. 

Since $H$ is full-rank as shown in theorem~\ref{thm:secure-capacity}, its sub-matrix $H_{\Gprime}$ will be full rank for any $\Gprime\subseteq[\Bs]$. Further, it has been shown in~\cite{Shah:11} that for PM-MBR codes each storage node stores $\al$ linearly independent symbols, thus, it follows that $G_e$ is full-rank. Therefore, to prove~\eqref{eq:subspace-condition}, we need to prove that the matrix $T$ is full-rank. We divide the proof into three cases: $k\geq 3$, $k = 2$, and $k=1$.

\textbf{Case 1:} $k\geq 3$. 

As in the proof of Theorem~\ref{thm:secure-capacity}, we show that, if $|\Gprime|\leq d+k-3$, it is always possible to  append $B-|\Gprime|-\alpha$ rows of appropriate types to $T$ in such a way that the resulting $B\times B$ matrix is non-singular. 
In the following, we present an algorithm which, for any given $H_{\Gprime}$ and node index $e\in[n]$, appends row vectors of appropriate types to $T = \twomatrix{H_{\Gprime}}{G_e}$ in such a way that successive decoding can be carried out. 

We need some notation to describe the algorithm. Let $\lamp_i$ be the number of encoding vectors of type $i$, $i\in[d]$, that are present in $H_{\Gprime}$. Notice that $\lamp_i \leq \theta_i$ $\forall i\in[d]$. 
Let $\lam_i$ denote the number of row vectors of type $i$, $i\in[d]$, that are present in $T$. Note that, for any $e\in[n]$, $G_e$ contains one row vector of each of the $d$ types. Thus, $\lam_i = \lamp_i + 1$, $\forall i\in[d]$. This implies that $\lam_i \leq \theta_i + 1$ $\forall i\in[d]$. Further, from~\eqref{eq:card-theta-PM-MBR}, we have that $\lam_i \in \{1,2\}$ for all group II types $i$ for $k+1\leq i\leq d$. Let $\gam_1$ and $\gam_2$ denote the number of equations that are required to decode the variables of group I and group II types, respectively, in a given iteration.
See algorithm~\ref{alg:decoding-T} on next page.

\begin{algorithm}[!t]
\caption{Appending rows to $T$ to form $T'$ and carrying out successive decoding for $Z=T'Y$ $(k\geq 3)$}
\label{alg:decoding-T}
\begin{algorithmic}[1]
\STATE Sort $\lam_j$ for $j\in[k]$, Let $\lams{{j_1}}\leq\cdots\leq\lams{{j_k}}$
\STATE Sort $\lam_j$ for $k+1\leq j\leq d$, Let $\lams{{j_{k+1}}}\leq\cdots\leq\lams{{j_d}}$
\STATE Find $L$ such that $\lams{j_{d-L+1}} = \lams{j_{d-L+2}} = \cdots = \lams{j_d} = 2$
\STATE \COMMENT{Notice that $0\leq L\leq d-k$}
\STATE {Set $\gam_1 = d$, $\gam_2 = k$}
\FOR{$p=k$ \TO $3$}
	\IF{$\lams{j_p} > \gam_1$}
		\STATE{Declare failure, Exit}
	\ELSE
		\STATE{Append $T$ with $\gam_1 - \lams{j_p}$ additional rows of type $p$ with non-zero coefficients as the rows of $\Psih$ that are not present in the $\lams{j_p}$ rows of type $p$}
		\STATE{Using the equations corresponding to the $\gam_1$ rows of type $p$, decode the un-decoded variables from $Y[\Ind{p}]$}
		\STATE{$\gam_1 = \gam_1 - 1$, $\gam_2 = \gam_2 - 1$}
	\ENDIF
\ENDFOR
\STATE \COMMENT{At this point, $\gam_1 = d - (k-2)$ and $\gam_2 = k - (k-2) = 2$}
\IF{$L > 0$}
	\STATE{Successively decode the remaining variables from $Y[\Ind{j_{d-L+i}}]$ for $i\in[L]$} \label{line-1}
	\STATE{$\gam_1 = \gam_1 - L$}
\ENDIF
\STATE \COMMENT{At this point, $\gam_1 = d - (k-2) - L$ and $\gam_2 = 2$}
\IF{$\lams{j_2} > \gam_1$}
	\STATE{Declare failure, exit} \label{last-if}
\ELSE
	\STATE{Append $T$ with $\gam_1 - \lams{j_2}$ additional rows of type $2$ with non-zero coefficients as the rows of $\Psih$ that are not present in the $\lams{j_2}$ rows of type $2$}
	\STATE{Decode the un-decoded variables from $Y[\Ind{2}]$} \label{line-2}
	\STATE{$\gam_1 = \gam_1 - 1$, $\gam_2 = \gam_2 - 1$}
	\STATE \COMMENT{At this point, $\gam_1 = d - (k-2) - L - 1$ and $\gam_2 = 2 - 1 = 1$}
	\STATE{Decode for the remaining symbols from $Y[\Ind{j_{k+1}}], Y[\Ind{j_{k+2}}], \ldots, Y[\Ind{j_{d-L}}]$} \label{line-3}
	\STATE{Append $T$ with a row of type $1$ with non-zero coefficients as the first row of $\Psih$}
	\STATE{Decode the un-decoded variable from $Y[\Ind{1}]$}
\ENDIF
\end{algorithmic}
\end{algorithm}

First, we prove the correctness of the algorithm.
\begin{claim}
\label{clm:correctness-2}
If algorithm~\ref{alg:decoding-T} does not report a failure, it finds a solution to $Z=T'Y$, where the construction of $T'$ is described in the algorithm.
\end{claim}
\begin{IEEEproof}
In the same way as in the proof of claim~\ref{clm:correctness-1}, it is easy to prove by induction that in the first {\bf for} loop, the algorithm decodes for $Y[\Ind{i}]$ for $3\leq i\leq k$. Since each pair of group I types has one index in common, $k-2$ elements of each of the remaining types are decoded at the end of the first {\bf for} loop. Note that there are two rows each of types $j_{d-L+1}$ through $j_d$. Since $k-2$ elements of each of them are already decoded, the remaining two elements are decoded at line~\ref{line-1}.  

At line~\ref{line-2}, all the remaining elements of $\Ind{2}$ will be decoded. At this point, there is only one un-decoded element each in types $j_{k+1}$ through $j_{d-L}$, which will be decoded at line~\ref{line-3}. Note that, at this point, all the types from 2 through $d$ have been decoded. Thus, there remains only one un-decoded element of type 1 which will be decoded as the final step. For successive decoding, we rely on the fact that for matrix $\Psit_e = \twomatrix{\Psih}{\Psi_e}$, any square sub-matrix is non-singular. Note that this condition holds by construction 1; e.g., when the matrix $\Psit$ is a Cauchy matrix.

Essentially, algorithm decodes the elements in all the $d$ types in the following order ($j_k$, $j_{k-1}$, $\ldots$, $j_3$), ($j_{d-L+1}$, $\ldots$, $j_{d}$), ($j_{k+1}$, $\ldots$, $j_{d-L}$), $j_2$, $j_1$, which covers all the $B$ elements.
\end{IEEEproof} 

Next, we prove that the algorithm~\ref{alg:decoding-T} does not declare a failure if the number rows in $H_{\Gprime}$ is bounded below a threshold.
\begin{claim}
\label{clm:success-for-bounded-G}
If $|\Gprime| \leq d+k-3$, then algorithm~\ref{alg:decoding-T} always succeeds. 
\end{claim}
\begin{IEEEproof}
The proof is by contradiction. Suppose $|\Gprime| \leq d+k-3 $, and the algorithm fails. 

{\it Case 1:} Algorithm fails on line 6, \ie, in the first iteration when $p=k$. This implies that $\lams{j_k} > \gam_1 = d$. However, we have
\begin{IEEEeqnarray}{rCl}
\lams{j_k} & = & \max_{1\leq l\leq k}\lam_{l} \nonumber\\
& \stackrel{(a)}{\leq} & \max_{1\leq l\leq k}\theta_l + 1 \nonumber\\ 
& \stackrel{(b)}{=} d \nonumber
\end{IEEEeqnarray}
where (a) follows from the fact that the number of rows of any type $i$ in $T$ is at most $\theta_i + 1$, \ie, $\lam_i \leq \theta_i + 1$ $\forall i\in[d]$, and (b) is due to~\eqref{eq:card-theta-PM-MBR}. This is a contradiction, and the algorithm cannot fail in the first iteration when $p = k$.

{\it Case 2:} Algorithm fails in the first {\bf for} loop during $i$-th iteration such that $2\leq i\leq k-2$. Note that this implies $k\geq 4$. Also, at the $i$-th iteration, we have $p = k - (i - 1)$. 

Now, as $\gam_1$ is reduced by 1 in each iteration, in $i$-th iteration we have $\gam_1 = d - (i - 1)$. Since the algorithm failed, it should be that $\lams{j_{k-i+1}} > d - (i - 1)$.  Then, we can write
\begin{IEEEeqnarray}{rCl}
\sum_{l=1}^{i}\lams{j_{k-l+1}} & \stackrel{(c)}{\geq} & i \lams{j_{k-i+1}} \nonumber\\
& \stackrel{(d)}{\geq} & i(d-i+2) \nonumber\\
\label{eq:lam-to-lamp}
\therefore\quad i + \sum_{l = 1}^{i} \lams{j_{k-l+1}}' & \geq & i(d-i+2) \\
\label{eq:lamp-final}
\therefore\quad \sum_{l = 1}^{i} \lams{j_{k-l+1}}' & \geq & i(d-i+1),
\end{IEEEeqnarray}
where (c) is due to $\lams{j_k} \geq \lams{j_{k-1}} \geq \cdots \geq \lams{j_{k-i+1}}$ and (d) is due to $\lams{j_{k-i+1}} > d - (i - 1)$. To get~\eqref{eq:lam-to-lamp}, we use $\lam_l = \lamp_l + 1$ $\forall l\in[d]$. 

First, note that $|\Gprime| = \sum_{l=1}^{d} \lamp_l$. Thus, $|\Gprime| \geq \sum_{l = 1}^{i} \lams{j_{k-l+1}}'$. Next, notice that $f(i) = i(d-i+2)$ is a concave function in $i$ and thus it will attain minimum over $2\leq i\leq k-2$ at one of its boundary points. Using these two observations along with~\eqref{eq:lamp-final}, we have
\begin{equation}
\label{eq:lamp-final-1}
|\Gprime| \geq \min \{2(d-1),(d-k+3)(k-2)\}.
\end{equation}
Now, we show that both of these boundary points result in contradiction. First, since $d \geq k$ for any regenerating code, clearly, $2(d-1) - (d+k-3) = d-k+1 > 0$, \ie, $2(d-1) > d-k+3$. Next, consider the second boundary point,
\begin{IEEEeqnarray}{rCl}
(d-k+3)(k-2) - (d+k-3) & \stackrel{(e)}{=} & (k-3)d - (k-2)^2 +1\nonumber\\
& \stackrel{(f)}{\geq} & (k-3)k - (k-2)^2 + 1 \nonumber\\
& = & k-3, \nonumber
\end{IEEEeqnarray}
where (e) follows from algebraic manipulations, and (f) follows because for any regenerating code $d\geq k$. Finally, since $k\geq 4$ in this case, we have $(d-k+2)(k-2) > d+k-3$. Therefore, we have $|\Gprime| \geq \min \{2(d-1),(d-k+3)(k-2)\} > d+k-3$, which is a contradiction. 

{\it Case 3:} Algorithm fails at line~\ref{last-if}, while considering type $j_2$. It is easy to see that at this point $\gam_1 = d - (k-2) - L$ and $\gam_2 = 1$. The failure implies that $\lams{j_2} \geq d - (k-2) -L +1$. Now, let us consider the total number of rows in $T$ corresponding to types that have been considered so far, as follows.
\begin{IEEEeqnarray}{rCl}
\sum_{l=2}^{k}\lams{j_l} + \sum_{m = d - L + 1}^{d} \lams{j_m} & \stackrel{(g)}{\geq} & \sum_{l=2}^{k}\lams{j_2}  + 2L \nonumber\\
\label{eq:total-rows-T}
& \stackrel{(h)}{\geq} & (k-1)(d-k-L+3) + 2L,
\end{IEEEeqnarray}
where (g) follows from $\lams{j_2} \leq \lams{j_3} \leq \cdots \leq \lams{j_k}$ and $\lams{j_{d - L + 1}} = \lams{j_{d - L + 2}} = \cdots = \lams{j_d} = 2$, and (h) follows from $\lams{j_2} \geq d - (k-2) -L +1$.
However, since $\lam_l = \lamp_l + 1$ for each $l\in[d]$, from~\eqref{eq:total-rows-T}, we can write
\begin{IEEEeqnarray}{rCl}
\sum_{l=2}^{k}\lams{j_l}' + \sum_{m = d - L + 1}^{d} \lams{j_m}' & \geq & (k-1)(d-k-L+3) - (k-1) + L\nonumber\\
\label{eq:bound-rows-HG}
& = & (k-1)(d-k-L+2) + L.
\end{IEEEeqnarray}
After some manipulations, it is straightforward to show that $(k-1)(d-k-L+2)+L - (d+k-3) = (d-k-L)(k-2) + 1$, which is strictly positive for $k\geq 3$ as $d\geq k$ and $0\leq L\leq d-k$. Therefore, we have 
\begin{equation}
\label{eq:total-rows-HG}
|\Gprime| = \sum_{l=1}^d\lams{j_l}' \geq \sum_{l=2}^{k}\lams{j_l}' + \sum_{m = d - L + 1}^{d} \lams{j_m}' \stackrel{(o)}{\geq} (k-1)(d-k-L+2) + L \stackrel{(r)}{>} d+k-3,
\end{equation}
where (o) follows from~\eqref{eq:bound-rows-HG} and (r) is proved in the previous point. However, this is a contradiction, which completes the proof for $k\geq 3$.

\end{IEEEproof}

\textbf{Case 2:} $k=2$.

We present the algorithm for successive decoding as follows.

\begin{algorithm}[!t]
\caption{Appending $T$ and carrying out successive decoding for $k = 2$}
\label{alg:decoding-T-k-2}
\begin{algorithmic}[1]
\STATE Sort $\lam_j$ for $k+1\leq j\leq d$, Let $\lams{{j_{k+1}}}\leq\cdots\leq\lams{{j_d}}$
\STATE Find $L$ such that $\lams{j_{d-L+1}} = \lams{j_{d-L+2}} = \cdots = \lams{j_d} = 2$
\STATE \COMMENT{Notice that $0\leq L\leq d-k$}
\STATE {Set $\gam_1 = d$, $\gam_2 = k = 2$}
\IF{$L > 0$}
	\FOR{$p=1$ \TO $L$}
		\STATE{Using the equations corresponding to the $\gam_2$ rows of type $d-L+p$, decode the variables indexed by $\Ind{d-L+p}$} \label{a2-line-1}
		\STATE{Set $\gam_1 = \gam_1 - 1$}
	\ENDFOR
\ENDIF
\STATE \COMMENT{At this point, $\gam_1 = d - L$}
\IF{$\lams{j_k} > \gam_1$}
	\STATE{Declare failure, exit} \label{a2-failure-point} 
\ELSE
	\STATE{Append $T$ with $\gam_1 - \lams{j_k}$ additional rows of type $k=2$ with non-zero coefficients as the rows of $\Psih$ that are not present in the $\lams{j_2}$ rows of type $2$}
	\STATE{Decode the un-decoded variables from $Y[\Ind{2}]$} \label{a2-line-2}
	\STATE{$\gam_1 = \gam_1 - 1$, $\gam_2 = \gam_2 - 1$}
	\STATE \COMMENT{At this point, $\gam_1 = d -  L - 1$ and $\gam_2 = 2 - 1 = 1$}
	\STATE{Decode for the remaining symbols from $Y[\Ind{j_{k+1}}], Y[\Ind{j_{k+2}}], \ldots, Y[\Ind{j_{d-L}}]$} \label{a2-line-3}
	\STATE{Append $T$ with a row of type $1$ with non-zero coefficients as the first row of $\Psih$}
	\STATE{Decode the un-decoded variable from $Y[\Ind{1}]$}
\ENDIF
\end{algorithmic}
\end{algorithm}

First, we prove the correctness of the algorithm.
\begin{claim}
\label{clm:correctness-3}
If algorithm~\ref{alg:decoding-T-k-2} does not report a failure, it finds a solution to $Z=T'Y$, where $T'$ is the matrix resulting after appending the rows to $T$ as described in the algorithm.
\end{claim}
\begin{IEEEproof}
As showed in the discussion before claim~\ref{clm:success-for-bounded-G}, notice that there are two elements each in $\Ind{l}$ for $k+1 \leq l\leq d$. For the $L$ types, $j_{d-L+1}$ through $j_d$, there are two rows present in $T$. Thus, all the variables from $\Ind{j_l}$ for $d-L+1\leq l\leq d$ are decoded at line~\ref{a2-line-1}.

Algorithm will decode all the un-decoded variables indexed by $\Ind{k=2}$ at line~\ref{a2-line-2}. Then, there remains only one un-decoded element of types $j_{k+1}$ through $j_{d-L}$, which will be decoded in line~\ref{a2-line-3}. At this point, there remains only single un-decoded variable from type 1, ant it is decoded as the final step. 

Essentially, algorithm covers all the $d$ types in the order ($j_{d-L+1}$, $\ldots$, $j_{d}$), ($j_{(k=2)}$), ($j_{k+1}$, $\ldots$, $j_{d-L+1}$), $j_1$, and decodes all the $B$ elements.
\end{IEEEproof}

\begin{claim}
\label{clm:success-for-bounded-G-k-2}
If $|\Gprime| \leq d+k-3 = d-1$, algorithm~\ref{alg:decoding-T-k-2} always succeeds. 
\end{claim}
\begin{IEEEproof}
The proof is by contradiction. Suppose $|\Gprime|$ and the algorithm fails. The only possibility of failure is line~\ref{a2-failure-point}. Consider the total number of rows in $T$ that have been considered till this point as follows.
\begin{IEEEeqnarray}{rCl}
\sum_{m = d - L + 1}^{d} \lams{j_m} + \lams{j_2} & \stackrel{(a)}{\geq} & (d - L + 1)  + 2L \nonumber\\
\label{eq:total-rows-T-2}
\therefore \quad \sum_{m = d - L + 1}^{d} \lams{j_m}' + L +\lams{j_2} + 1 & \geq & d+L+1, \\
\label{eq:total-rows-T-3}
\therefore \quad \sum_{m = d - L + 1}^{d} \lams{j_m}'  +\lams{j_2} & \geq & d, 
\end{IEEEeqnarray}
where (a) follows because failure implies $\lams{j_2} > d-L$. To get~\eqref{eq:total-rows-T-2}, we use $\lams{j_l} = \lams{j_l}' + 1$ for each $l\in[d]$. Now, we can write
\begin{IEEEeqnarray}{rCl}
|\Gprime| & = & \sum_{l=1}^{d} \lams{j_l}' \nonumber\\
& \geq & \sum_{m = d - L + 1}^{d} \lams{j_m}'  +\lams{j_2} \nonumber\\
& \stackrel{(b)}{\geq} & d,
\end{IEEEeqnarray}
where (b) is due to~\eqref{eq:total-rows-T-3}. However, $|\Gprime| \geq d$ is a contradiction and the proof follows.
\end{IEEEproof}

\textbf{Case 3:} $k=1$.

Notice that for $k=1$, Eve gets the same degrees of freedom as a data collector (which accesses to $k$ nodes to recover the file). Therefore, it is not possible to achieve any form of security since, similar to the data collector, Eve can also decode the complete file 

\textbf{Example:} Consider one possible matrix $T$ as shown in~\eqref{eq:T-ex}.
First, append one row of type 4 and decode for the variables indexed by $\Ind{4} = \{4, 7, 9\}$. Then, add one row of type 3 and decode for the variables indexed by $\Ind{3}\setminus((\Ind{4}\cap\Ind{3})) = \{3, 6, 8\}$. Using the two rows of type 2, decode for the variables indexed by $\Ind{2}\setminus((\Ind{4}\cap\Ind{2})\cup(\Ind{3}\cap\Ind{2})) = \{2, 5\}$. Finally, using a row of type 1, decode for the variable indexed by $\Ind{1}\setminus((\Ind{1}\cap\Ind{4})\cup(\Ind{1}\cap\Ind{3})\cup(\Ind{1}\cap\Ind{2})) = \{1\}$.  The non-zero coefficients of the appended rows are specified by the appropriate rows of $\Psih$, and the successive decoding is feasible due to the property that any square sub-matrix of the Cauchy matrix $\Psit = \twomatrix{\Psi}{\Psih}$ is non-singular.

\begin{equation}
\label{eq:T-ex}
T = \twomatrix{H_{\Gprime}}{G_{e}} = 
\begin{bmatrix}
0 & \Psih(2,1) & 0 & 0 & \Psih(2,2) & \Psih(2,3) & \Psih(2,4) & 0 & 0\\
0 & 0 & \Psih(3,1) & 0 & 0 & \Psih(3,2) & 0 & \Psih(3,3) & \Psih(3,4)\\
0 & 0 & 0 & \Psih(1,1) & 0 & 0 & \Psih(1,2) & 0 & \Psih(1,3)\\
\Psi(e,1) & \Psi(e,2) & \Psi(e,3) & \Psi(e,4) & 0 & 0 & 0 & 0 & 0\\
0 & \Psi(e,1) & 0 & 0 & \Psi(e,2) & \Psi(e,3) & \Psi(e,4) & 0 & 0\\
0 & 0 & \Psi(e,1) & 0 & 0 & \Psi(e,2) & 0 & \Psi(e,3) & \Psi(e,4)\\
0 & 0 & 0 & \Psi(e,1) & 0 & 0 & \Psi(e,2) & 0 & \Psi(e,3)
\end{bmatrix}
\end{equation}





\bibliographystyle{IEEEtran}
\bibliography{IEEEabrv,Regen_bib_1}
%
%
%

\end{document}